# Process Mining Meets Visual Analytics: The Case of Conformance Checking

Jana-Rebecca Rehse
University of Mannheim
rehse@uni-mannheim.de

Luise Pufahl
Technische Universitaet Berlin
luise.pufahl@tu-berlin.de

Michael Grohs
University of Mannheim
michael.grohs@uni-mannheim.de

Lisa-Marie Klein
University of Mannheim
liklein@mail.uni-mannheim.de

## Abstract

*Conformance checking is a major function of process mining, which allows organizations to identify and alleviate potential deviations from the intended process behavior. To fully leverage its benefits, it is important that conformance checking results are visualized in a way that is approachable and understandable for non-expert users. However, the visualization of conformance checking results has so far not been widely considered in research. Therefore, the goal of this paper is to develop an understanding of how conformance checking results are visualized by process mining tools to provide a foundation for further research on this topic. We conduct a systematic study, where we analyze the visualization capabilities of nine academic and seven commercial tools by means of a visual analytics framework. In this study, we find that the "Why?" aspect of conformance checking visualization seems already be well-defined, but the "What?" and "How?" aspects require future research.*

## 1. Introduction

Process mining refers to a family of data analysis techniques aimed at extracting knowledge about business processes from event logs that capture the execution of those processes in information systems. One of the main functions of process mining is conformance checking, which aims to compare the to-be business process, described by a process model, with the as-is process execution, as captured in an event log (Carmona et al., 2018). Therefore, individual process executions are matched to the event log to find situations where the execution of the process does not follow the process model, e.g., because an activity is skipped. Applying conformance checking techniques enables organizations to check whether their process execution complies with the intended process behaviour (Munoz-Gama, 2016). They also serve as an automated tool to keep process models up to date and identify optimization potentials (Carmona et al., 2018).

These capabilities make conformance checking highly relevant for organizations from many different domains (Emamjome et al., 2019). To increase its applicability, the computational efficiency of conformance checking has been considerably improved (Dunzer et al., 2019). It has also been extended beyond a pure control flow view to consider, e.g., resources and data (Gall & Rinderle-Ma, 2017; Knuplesch et al., 2017). However, powerful algorithms are not sufficient to fully leverage the benefits of conformance checking in practice. Another relevant aspect is the visualization of conformance checking results (Garcia-Banuelos et al., 2017; Gschwandtner, 2017). When analyzing large amounts of complex data, humans depend on high-quality visualizations to quickly access new information, efficiently draw conclusions, and eventually make better decisions for their organization (Keim et al., 2008). Despite its practical relevance, the visualization of conformance checking results has so far not been widely considered in research (Garcia-Banuelos et al., 2017; Gschwandtner, 2017; Klinkmüller et al., 2019).

The process mining community has developed many software tools for practical process analysis (FAU, 2020). Those include academic research frameworks, such as ProM[1], and commercial tools, such as SAP Signavio[2]. For both academic and commercial tool providers, it is vitally important to communicate analysis results with appropriate visualization techniques to convince analysts of their tool's value. A recent study among practitioners found that conformance checking is the most critical feature for selecting a process mining tool (FAU, 2021), which makes this aspect particularly relevant for conformance checking. Nowadays, most tools include conformance checking techniques (FAU, 2020), which are increasingly used to compute and visualize results in application projects (Emamjome et al., 2019).

---
[1] http://www.promtools.org/
[2] https://www.signavio.com/process-mining/

To summarize, there is an ongoing development of visualization approaches for conformance checking in process mining tools, which has so far not received much attention in research, despite its practical relevance. Therefore, this paper aims to develop an understanding of how process mining tools visualize conformance checking results to provide a foundation for future research on this topic. To this end, we conduct a systematic study of nine academic and seven commercial tools that provide functionalities for visualizing conformance checking results. In this study, we apply a methodical framework from the field of visual analytics (Munzner, 2014), the research discipline concerned with designing interactive visual interfaces to support analytical reasoning (Keim et al., 2008). This framework allows us to systematically assess the "What?", "How?", and "Why?" of visualization. Our findings can guide future research in both process mining and visual analytics to advance the discipline and further improve the practical applicability of conformance checking (Gschwandtner, 2017; Klinkmüller et al., 2019).

In the following, we provide the necessary background on visual analytics and conformance checking in Sect. 2 and report on related work in Sect. 3. Sect. 4 outlines the method that we used to analyze visualizations in process mining tools. The results of this analysis are presented in Sect. 5. We present main insights and future research opportunities in Sect. 6, before concluding the paper in Sect. 7.

## 2. Background

Our paper analyzes conformance checking tools through a visual analytics lens. In the following, we provide the necessary background on both fields.

### 2.1. Visual Analytics

Visualization refers to communicating data by means of interactive interfaces (Keim et al., 2008). Its goals are (i) presenting analysis results, (ii) enabling the confirmation of assumed relations in the data, and (iii) allowing for an undirected exploration of the data. Visual analytics is the science of analytical reasoning supported by interactive visual interfaces. It extends visualization to focus on the interplay between visualization, data analysis, and interaction methods. As such, it is not an independent research field but is closely intertwined with the covered domains.

Visualization aims to abstract knowledge from large amounts of data (Keim et al., 2008). To systematize this knowledge abstraction, the visual analytics framework, shown in Fig. 1, divides each visualization into three aspects: "What?", "Why?", and "How?" (Munzner, 2014). "What?" refers to the visualized data, including its items, nodes or links, and attributes (Raimbaud et al., 2019). "Why?" refers to the task that is performed. It can typically be articulated in two words: one action (verb) and one target (noun), such as "summarize distribution" (Raimbaud et al., 2019). "How?" refers to the elements used to visualize the data, called visualization idioms. They consist of visual channels like color or size and visual representations like lines or areas (Raimbaud et al., 2019).

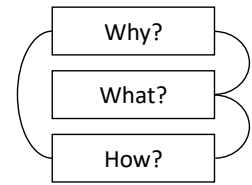

Figure 1: Visual Analytics Framework (Munzner, 2014)

Based on this framework and others, the visual analytics community has built multiple knowledge bases that collect and structure available visualization idioms. One is the Data Visualisation Catalogue[3]. This catalog, which combines and integrates visualization idioms from several other online sources, is supposed to serve as a reference tool for data visualization. According to its citations on Google Scholar, it is frequently cited by researchers from visual analytics and other domains.

### 2.2. Conformance Checking

Conformance checking relates the intended behavior of a process, as depicted by a process model, with the actual behavior of a process, as recorded in an event log (Carmona et al., 2018). Process models describe a set of activities and the order in which they should be executed to reach a certain process goal (Weske, 2019). Researchers often model processes as Petri nets, but the de-facto industry standard is Business Process Model and Notation (BPMN). Commercial process mining tools often model processes with simple directly-follows graph (DFG) (Leemans et al., 2019). Event logs are collections of event sequences describing the activities that have occurred during the execution of a business processes (Carmona et al., 2018). Each event is assigned to a specific business case, such that all events of a case form a trace that represents its execution. A set of traces with the same events in the same order is called a trace variant. For each variant, the frequency and the percentage of occurrence in an event log can be calculated. Given a process model and an event log, a conformance checking technique calculates a fitness or conformance value, which measures the amount of log behavior that conforms to the model.

---
[3] https://datavizcatalogue.com/

The current state-of-the-art conformance checking technique are trace-level alignments (Carmona et al., 2018). For each activity in a trace, we search for a corresponding move in the process model (synchronous move). If the log activity is not enabled in the model, we generate a log move. If another model activity must be executed to enable the log activity, we generate a model move. Log and model moves are penalized by a cost function, which allows finding the optimal alignment for each trace. The log fitness is calculated as the average trace-level fitness, which is computed by dividing the cost of the optimal alignment by the cost-of the worst-case scenario and deducting the result from 1.

## 3. Related Work

So far, the visualization of conformance checking results has not been widely considered in research (Garcia-Banuelos et al., 2017; Klinkmüller et al., 2019). The usual feedback that conformance checking techniques provide is an overall fitness measure between log and model (Dunzer et al., 2019). If more detail is required, deviations are either illustrated individually per trace (de Leoni et al., 2015) or as model-level patterns (Gerke & Tamm, 2009). However, these types of feedback typically do not provide enough details for organizations to draw meaningful conclusions about the problems of their process. Klinkmüller et al., 2019 addressed this issue in their analysis of information needs in process mining. After analyzing 71 process mining reports, they found that conformance checking results are mainly conveyed by means of bar charts, tables, or process models but lack problem-specific presentation modes. The authors promote future research on domain-specific presentations to improve the interpretability of conformance checking results.

To address this issue, a few authors have recently published new conformance checking techniques explicitly focusing on visualization or other forms of result communication. Garcia-Banuelos et al., 2017 suggest an approach that provides a textual representation of behavior deviations to the users. This approach was empirically evaluated in comparison to conventional trace alignments. It was shown that professionals preferred text, whereas academics preferred trace alignments. Additionally, respondents with less experience in Petri nets had stronger preferences toward a textual representation.

Gall and Rinderle-Ma, 2017 introduced Instance Spanning Compliance States as an approach that extends BPMN to visualize constraints between different processes or instances. In a user study, they found that the interviewed experts welcomed the use of color for an overview of the constraint states and that they favored text over symbols for the possibility of expressing detailed information concisely (Gall & Rinderle-Ma, 2021). Other visualization-focused conformance checking approaches include the extended Compliance Rule Graph, a rule-based modelling language that includes resources and data objects in addition to control flow (Knuplesch et al., 2017), and the Probabilistic Inductive Miner, which can be used to construct visual alignments between the discovered model and to the to-be model (Brons et al., 2021).

The visual analytics field has also explored process conformance or compliance as a potential application field (Gschwandtner, 2017). It mainly focuses on healthcare processes, which typically follow so-called clinical guidelines, i.e., recommendations on how to treat a patient cohort. Bodesinsky et al., 2013 present an interactive visual approach to analyze the compliance of a single care process instance with these guidelines. It can be used in real-time to support clinical staff in caring for a patient and in retrospect to support experts in examining care quality and the guideline itself (Gschwandtner, 2017). Based on this, Basole et al., 2015 design another domain-specific visualization approach that allows to analyze many care process instances at the same time. Beyond the conformance checking domain, Yeshchenko et al., 2021 have employed a visual analytics approach for detecting process drifts in event log data. To conclude, both process mining and visual analytics researchers have recognized and acknowledged the mutual benefits that would arise from combining both fields, but this combination has not been studied systematically.

## 4. Research Method

In the following, we describe the design of our research study, including how the tools were selected and how the visualizations were analyzed by means of the visual analytics framework.

### 4.1. Tool Selection

**Academic tools.** To identify relevant academic tools, we conducted a structured literature review in academic papers (Kitchenham, 2004) to find the tools that they used to visualize conformance checking. We combined the term "conformance checking" with at least one of the terms "visual*", "graph", or "display" as potential textual descriptions of visualizations. Because "compliance checking" (combined with "process mining") is sometimes used as a synonym for "conformance checking", we repeated the search terms in that combination. We selected five scientific databases (AIS Electronic Library, Academic Search Premier, IEEE Xplore, ScienceDirect, and SpringerLink) and filtered for business and computer science. To balance relevance and size of the search space, only the 250 most relevant results per term and database were considered. In a second step, we assessed the identified literature for relevance concerning our research question. Therefore, each paper was searched for visualization of conformance checking results beyond the examples in Sect. 2. If only a textual description was provided or the process of computation was depicted, the respective paper was not classified as relevant. Applying these criteria and removing redundant contributions resulted in 42 papers. From those, we conducted a forward and a backward search to identify relevant more recent or prior contributions. We ended up with 68 papers.[4]

These papers were examined regarding the tools that they used for visualization, leading to nine academic tools in total: (Data-Aware) Declare Replayer (Bergami et al., 2021; de Leoni et al., 2015), Declare Analyzer (Burattin et al., 2016), Conformance Checker (Rozinat & van der Aalst, 2008; vanden Broucke et al., 2013), Directly Follows Visual Miner (Leemans et al., 2019), Inter-Level Replayer (Alizadeh et al., 2018), Planning-Based Alignment of Event Logs and Petri Nets (Lanciano, 2018), Replay a Log on Petri Net for Conformance Analysis (Estanol et al., 2019), Apromore Compare (Armas Cervantes et al., 2017), and Probabilistic Inductive Miner (Brons et al., 2021). Except for the last two, all of them are ProM plug-ins.

**Commercial tools.** To select the commercial process mining tools, we referred to a recent study (FAU, 2020), which includes a list of 17 commercial process mining tools. From this list, we pre-selected the ten tools that support conformance checking. We requested an academic license for the conformance checking components, which we received for seven: ARIS PM (ARIS), Celonis, LanaLabs (LanaL.), Mehrwerk Process Mining (MPM), myInvenio IBM (myInv.), PAFnow[5], and SAP PI by Signavio (SAP).

In our following analysis, we focus on the visualization only. We do not report on the type of conformance checking techniques that the tools use because this information is confidential for the commercial tools and not relevant for the visualizations. We also do not evaluate the correctness of the results.

### 4.2. Tool Analysis

To evaluate the tool's conformance checking visualizations, we used a publicly available event log on road traffic fines.[6] Using ProM, we discovered one imperative and one declarative process model with 20% noise filtering, and transformed the imperative model into a BPMN model, which we could upload in five of the seven commercial tools. In ARIS and PAFnow, the process model had to be created in the tool itself. For the academic tools, we could upload the model in the required format (declarative or imperative) for seven of the nine academic tools. However, for some tools, we had to significantly reduce the event log size to execute the conformance checking algorithms successfully. One tool (Alizadeh et al., 2018) could not be executed properly and another (Brons et al., 2021) was not publicly accessible. For those, we based our analysis on publicly available screenshots and documentations.

For each tool, we applied the available conformance checking techniques to the model and the event log and took screenshots to document how their results were visualized.[7] The visualizations were analyzed in terms of the visual analytics framework (see Fig. 1). Therefore, we inspected each visualization separately to identify the respective "What?", "Why?", and "How?". More specifically, we investigated what data was visualized (e.g., conformance levels, or deviation distributions, etc.), why it was visualized (e.g., quantify conformance level, or compare conformance distribution, etc.), and how the data was portrayed (e.g., line diagrams, or process diagrams, etc.). For each aspect, we applied a coding process, where the visualization was classified into one category. This ensured comparability of results. Following the guidelines for qualitative content analysis (Mayring, 2000), we inductively defined our categories for the "What?" and "Why?" aspects because we intended to capture the specificities of the domain as precisely as possible. These categories must be mutually exclusive and exhaustive, such that they allow for the unambiguous classification of all visualizations.

---

[4]A detailed list of all 68 relevant papers can be found at https://doi.org/10.6084/m9.figshare.17061266.

[5]In the time between the analysis and the paper submission, PAFNow was acquired by Celonis. However, the tools are not integrated (yet), so we decided to keep PAFNow as a separate tool.

[6]https://data.4tu.nl/articles/dataset/Road_Traffic_Fine_Management_Process/12683249

[7]The materials are available from the authors upon request.

We inferred the "Why?" categories from the first commercial tool and found that they could be applied to all others, with some slight modification of the descriptions. The "What?" categories were derived over the course of our analysis. Because the "What?" was always clearly named by the tools, the categories did not undergo any modifications and it was always clear when to add a new one. For the "How?" categories, we used the visualization idioms from the DataViz catalogue[8]. When we came across idioms not included in this catalogue, we added them as a new category. The added categories were either standard visualization idioms, such as numbers, tables, and text, or domain-specific visualizations, such as process models, such that their definition was always clear. If we observed that color was used to highlight specific aspects of conformance, e.g., conformant vs. non-conformant cases, we noted that a colored idiom was used and, where necessary, explained details textually.

## 5. Conformance Checking Visualizations

Concerning the "Why?" aspect, we inspected how the visualizations were structured in the user interfaces of the tools and found four high-level categories: (1) quantify conformance, (2) break-down and compare conformance, (3) localize and show deviations, (4) explain and diagnose deviations. In the following, we present our results subdivided by these categories. Using two-dimensional bubble plots, we show "What?" data (center) is visualized "How?" in the academic tools (left) and commercial tools (right). The darker colored bubbles in the respective outer column represent the sums of the rows To illustrate some of our findings, we include exemplary screenshots from the analyzed tools. Those are mainly meant to provide a high-level overview, so readability is not necessary.

### 5.1. Quantify Conformance

Conformance metrics provide a first impression and an important high-level indication about the conformance state of a business process. As seen in Fig. 2, the most-shown metric is average fitness. Additionally, minimum and maximum fitness, as well as the alignment moves (sum of simultaneous, model, and log moves) in the optimal alignment are shown by the academic tools, using only numbers.

In contrast to the academic tools, the commercial tools show a broader range of high-level metrics and use a few more visualization idioms. As shown in Fig. 2, four out of the seven commercial tools present the average fitness, and two also provide the minimum and maximum trace fitness. These measures are either represented with numbers, or even colored numbers that indicate whether the fitness is in the normal range or not. Three tools portray the conformance rate (i.e., the number of conformant cases in relation to all cases). Furthermore, five commercial tools provide the total number of (non-)conformant cases, and three even show the performance of conformant vs. non-conformant cases. Some tools use colors for these metrics to highlight conformant traces in green and non-conformant cases in red. ARIS uses a donut chart to give users an overview of conformant vs. non-conformant cases. Further, the total number of observed deviations is given by two commercial tools.

[8]https://datavizcatalogue.com/

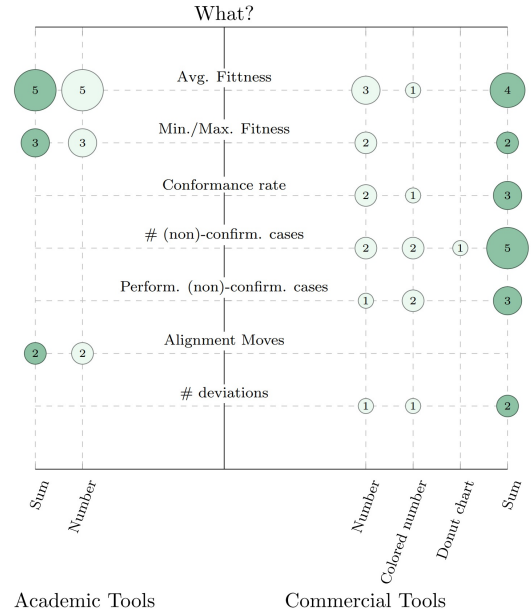

Figure 2: Visualizations for "Quantify conformance"

### 5.2. Break Down and Compare Conformance

To provide more insights into key metrics, some of the commercial tools offer additional diagrams, where the average fitness or the number of deviations are broken down further to study their distribution over multiple perspectives, as given in Fig. 3. No academic tool provides any visualizations in this category. LanaL. and MPM show how many traces reach a specific fitness range (up to 10%, ..., up to 100%) as frequency distribution in a bar chart. ARIS shows the fitness value per variant as a bar chart and the temporal fitness distribution as a scatter plot, shown in Fig. 4. Two other tools show the fitness over time periods, e.g., months, as an area graph or a bar chart. Additionally, MPM visualizes a geographical distribution of the fitness for different locations as a bubble map (a world map with bubbles of different sizes to represent the fitness values). MPM also provides a distribution of the fitness values for selected case attributes as a bar chart and in a table.

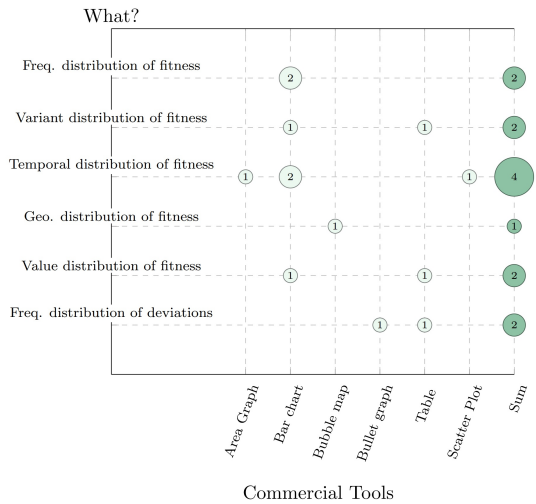

Figure 3: Visualization for "Break Down Conformance"

Regarding the number of deviations, ARIS includes a bullet graph that presents the number of deviations in predefined categories, such as *undesired activity* or *unexpected start/end activity*. Similarly, MPM presents a table that breaks down the deviations into *skipped activities*, *changed ordering*, and *skipped ordering* and provides the numbers per trace variant. We observed that MPM provides a broad range of visualization idioms to analyze the distribution of fitness and deviations over time, per case attribution, over locations, etc., such that the user can try to find hypotheses that could explain a lower fitness.

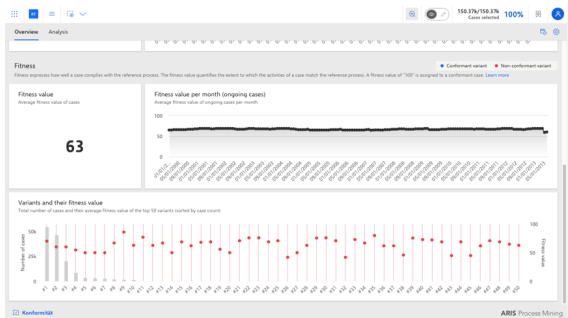

Figure 4: Screenshot from ARIS.

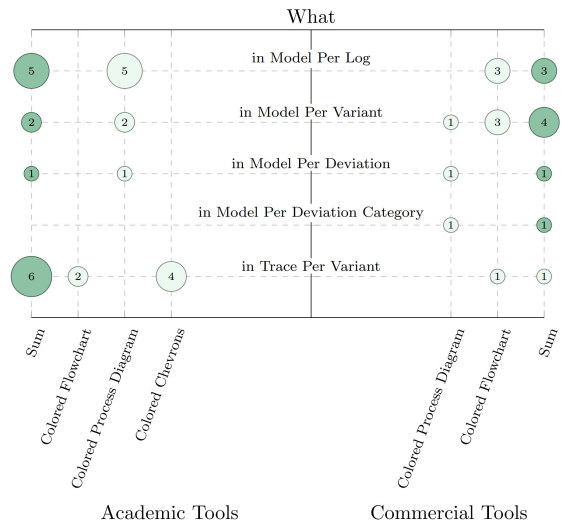

Figure 5: Visualization for "Localize Deviations"

### 5.3. Localize and Show Deviations

Several academic and commercial tools use domain-specific visualizations for deviations from the intended process design. They include process diagrams that adhere to a modeling language, like Petri nets or BPMN, more general flowcharts that do not adhere to a modeling language, and colored chevron diagrams (Carmona et al., 2018) that represent deviations in alignments. In the academic tools, we see two groups of approaches, as shown in Fig. 5. The first one uses diagrams of a specific modeling language to present the deviations on different levels of granularity. Most often, the deviations are shown per log. Sometimes, the deviations of one variant and each deviation individually are shown additionally. For example, the Apromore plugin uses a visualization per deviation in a BPMN diagram. As given in Fig. 6, the user can select a deviation, for which the deviating behavior is then visualized in red on top of the designed process model in black. The second group shows the deviation in the variant. This group uses flowcharts and chevrons to show the alignments of all activities in the trace. Note that all three idioms include colors to present the conformance checking results.

Some commercial tools show the deviations per category abstractly, some show deviations in the model per variant, and some show all deviations in one model, as summarized in Fig. 5. LanaL. visualizes the deviations as different categories (additional work, activity skipped, order switched, alternative paths) in a BPMN diagram, as shown in Fig. 7. The red color is used to visualize deviations from the process diagram. It gives the user a high-level overview, where they can click on the deviation symbols and learn about details, e.g., how often the deviation occurred.

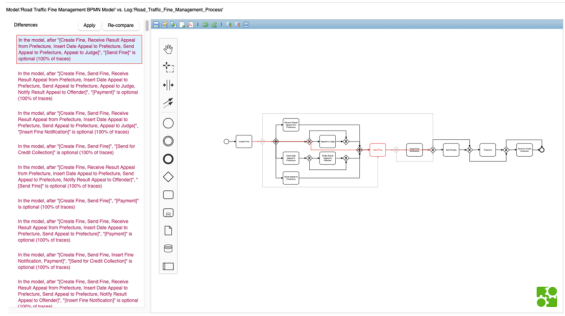

Figure 6: Screenshot from Apromore

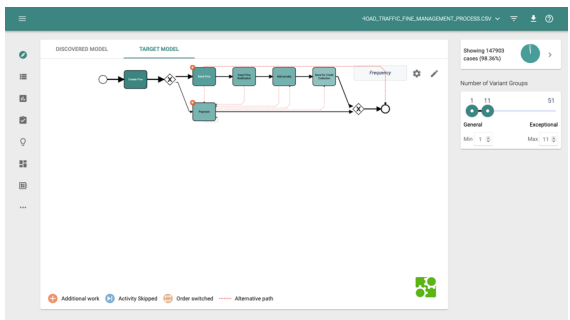

Figure 7: Screenshot from LanaLabs

Most commercial tools also analyze the compliance or deviation of the individual process variants. The user can select a variant and see how it deviates from the to-be process, as indicated by different colors. Three tools also allow showing all deviations and the designed process in one flow chart with color-coding, resulting in a highly complex visualization.

### 5.4. Explain and Diagnose Deviations

In addition to deviations in a model, we found that some academic and almost all commercial tools present details on the deviations, such as a textual description, the respective frequency, or a root cause analysis, as shown in Fig. 8. The academic tools utilize only text and numbers to present those details. In Sect. 3 we mentioned one empirical study that found textual descriptions of process deviations to be perceived as simpler than alignments (Garcia-Banuelos et al., 2017). The corresponding academic tool (Apromore Compare) is the only one that uses text to explain the deviation briefly. Another academic tool (Declare Analyzer)

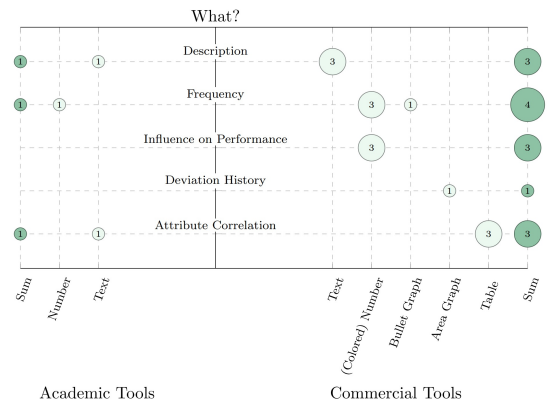

Figure 8: Visualizations for "Explain Deviations"

shows the attributes corresponding to deviations.

The commercial tools provide more details on the deviations and apply a broader range of visualization idioms. Three tools describe the deviation in a small text; most tools visualize the frequency of a deviation as a number. Three, including Celonis in Fig. 9, even report on the influence of a deviation on the performance as a (colored) number. Celonis also uses an area graph to show when and how often the violation has occurred in the past. Three tools provide the results of an attribute correlation, showing case attributes that often correlate with the traces including this deviation. MPM does the correlation analysis on all deviations, whereas Celonis and LanaL. provide the attribute correlation per deviation.

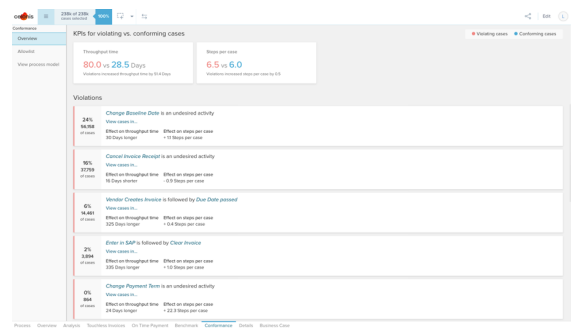

Figure 9: Screenshot from Celonis

Additional to the visualization of deviations' details, some commercial tools also provide the option to exclude identified deviations manually. This possibility has the advantage that the business analyst can manually correct missing aspects in the process model or wrong conformance checking results.

## 6. Potentials for Future Research

From our analysis, we can derive three main insights about current conformance checking visualizations:

**"Why?" Commercial tools provide a step-by-step analysis pipeline.** Regarding the "Why?", we found that commercial tools are more advanced than academic ones. They provide users with an analysis pipeline, where each step offers more fine-granular information. First, the overall process conformance is quantified by high-level metrics. Next, the process is broken down along multiple dimensions to show the distribution of conformance values and allow for a more detailed comparison. Afterward, the conformance deviations are localized in a process model to provide details about how the observed behavior did or did not conform with the prescribed process. Finally, the tools elaborate on the individual deviations, including potential root causes. The academic tools do not cover all steps of this pipeline but focus on the first and third one. None of them allows for breaking down and comparing conformance values, and only some provide visualizations explaining and diagnosing deviations.

**"What?" Few scientific insights on information needs and user preferences.** Considering the "What?", there are slight differences between academic and commercial tools. The commercial tools offer a few more options, such as showing deviations in a model per deviation category. However, none of the tools appear to have based their visualization choices on comprehensive (published) user studies. When analyzing related work, we only found one study that examined potential information needs that process mining tools should address (Klinkmüller et al., 2019), and a few contributions that examined potential users' preferences when being presented conformance checking results (Bodesinsky et al., 2013; Gall & Rinderle-Ma, 2021; Garcia-Banuelos et al., 2017). Only one of them (Garcia-Banuelos et al., 2017) related to one of the academic tools (Armas Cervantes et al., 2017) we examined. We did not observe any research or white papers that discuss a systematic user involvement in the developments of the "What?". There, we see a research potential for the future. In contrast to "Why?", we also observed that the "What?" differs significantly between the tools. We can conclude that there is no industry-wide agreement on operationalizing the individual steps.

**"How?" Tools mainly rely on standard visualization idioms.** Concerning the "How?", most used visualization idioms are standard: numbers, text, and tables. The slightly more advanced commercial tools offer a few additional idioms, such as bullet or area graphs. Still, tools mostly rely on standard visualization idioms found in the (domain-independent) DataViz catalog. An interesting insight, however, is that most tools enriched the (standard) idioms by using color, for example, to distinguish between conformant and non-conformant cases. We only found two domain-specific visualizations: colored chevron diagrams for the visualization of alignments and process modeling languages. Most academic tools use Petri Nets. Some commercial tools use BPMN, whereas the others revert to simple flow charts or DFGs.

Building on these insights, we propose a framework for future research on conformance checking visualization in Fig. 10. It instantiates the visual analytics framework to the conformance checking domain. We break the "Why?" aspect down into four steps: (1) quantify, (2) break-down and compare, (3) localize and show, (4) explain and diagnose. Although identified inductively, these steps represent an industry best practice to which the academic tools could easily be matched. We found that the main research potentials for conformance checking visualization concern the "What?" and "How?" aspects. Regarding the "What?", there are only very few studies that examine which data users need or prefer to be presented when applying conformance checking. We conclude that a systematic assessment of potential users' information needs and preferences is the main research potential for this aspect. This could be done by means of qualitative user interviews, a quantitative survey among practitioners, or observation and in-depth analysis of best practices in conformance checking. Regarding the "How?", we saw that there are few idioms for conformance checking visualization that went beyond basic numbers, texts, or tables. In analogy to the "What?", we conclude that a systematic assessment of potentially useful visualization idioms constitutes the main research potential for this aspect. It would entail a (1) structured analysis of existing visualization idioms and (2) the potential development and evaluation of new visualization idioms to find the most effective way to fulfill the needs and preferences identified in the previous aspect.

We intend the framework to be used from top to bottom: Researchers should select a "Why?" category and then systematically assess the "What?" for this category, identifying which data users need or want to see for their respective visualization purpose. Only after determining the "What?", they should address the "How?" by examining which visualization idioms would be best suited. Individually researching both aspects for all "Why?" categories might be time-intensive, but potentially beneficial to both process mining and visual analytics, hence increasing the applicability and relevance of conformance checking.

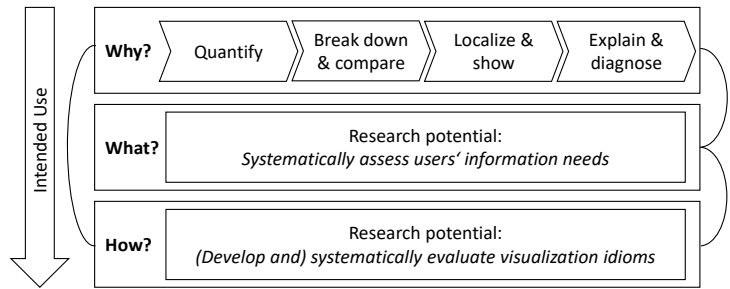

Figure 10: Framework for future research on conformance checking visualization

## 7. Discussion & Conclusion

This paper aimed to develop an understanding of how process mining tools visualize conformance checking results to provide a foundation for further research. We conducted a systematic study where we analyzed nine academic and seven commercial tools through the lens of a visual analytics framework. We found that the "Why?" aspect of conformance checking visualization is relatively well-defined: Visualizations could be categorized into one of four steps. Concerning the "What?" aspect, there were considerable differences between the tools, which might be explained by a lack of systematic studies on the topic. The "How?" was mainly based on standard visualization idioms. We hence concluded that the two main research potentials in conformance checking visualization are (1) a systematic assessment of users' information needs and preferences and (2) the potential development and systematic evaluation of suitable visualization idioms for these needs and preferences.

Our findings are subject to multiple threats to validity. First, our tool analysis may be incomplete. For the academic tools, a text-based search for visualizations might not identify all relevant results. Moreover, there might be newer academic tools that have not yet been used in scientific papers or demo tools that are not listed in the scientific databases. One option to address this on future work is to also analyze demo proceedings of major process mining conferences. For the commercial tools, the original list is most likely complete because it is based on a recent study by other researchers (FAU, 2020). Still, our findings might be limited because we only analyzed freely available academic versions.

Other threats concern our analysis approach. Particularly for the "Why?" and "What?" aspects, we defined the categories inductively without further guidance. Even though we carefully refined and documented the individual categories and had an independent second researcher double-check the coding of the individual tools, we cannot rule out that potential mistakes or inconsistencies occurred during the coding process. This also applies to the sources of our analysis approach. Because we could not access all academic tools, two of them were only analyzed via screenshots from different documentation sources. Although those sources appeared to be rather comprehensive, we cannot rule out that we missed some visualizations that were available in the tool but not contained in the documentation. However, we do not expect these two issues to impact the overall results majorly.

The final threat concerns comparing academic and commercial tools, which was not the main objective of our research but an inevitable consequence. We acknowledge that this comparison is not fair, given that academic and commercial tools might have different purposes and target groups and are developed under different circumstances. However, we argue that both types of tools are essential for the practical advancement of conformance checking. Given that academic tools are often freely available, practitioners might use them to test the capabilities of conformance checking. Hence, adequate visualization of conformance checking results in academic tools might pave the way for increased practical adoption of conformance checking.